\documentclass[aps,showpacs,pre,twocolumn,reprint]{revtex4-1}
\usepackage{graphicx,bm,amsmath,amssymb,natbib,url,epsfig}
\usepackage{dcolumn}
\usepackage{hyperref}

\usepackage{color}

\begin{document}

\title{Environmental coupling in ecosystems: From oscillation
  quenching to rhythmogenesis}

\author{Ramesh Arumugam}

\author{Partha Sharathi Dutta}
\thanks{Corresponding author}
\email{parthasharathi@iitrpr.ac.in}
\affiliation{Department of Mathematics, Indian Institute of Technology
  Ropar, Punjab--140 001,~India.}
\author{Tanmoy Banerjee} \email{tbanerjee@phys.buruniv.ac.in}
\affiliation{Department of Physics, University of Burdwan, West
  Bengal--713 104, India.}

\received{:to be included by reviewer}
\date{\today}

\begin{abstract}
How do landscape fragmentation affects ecosystems diversity and
stability is an important and complex question in ecology with no
simple answer, as spatially separated habitats where species live are
highly dynamic rather than just static.  Taking into account the
species dispersal among nearby connected habitats (or patches) through
a common dynamic environment, we model the consumer--resource
interactions with ring type coupled network. Characterizing the
dynamics of consumer--resource interactions in a coupled ecological
system with three fundamental mechanisms: such as the interaction
within the patch, the interaction between the patch and the
interaction through a common dynamic environment, we report the
occurrence of various atypical collective behaviors.  We show that,
the interplay between dynamic environment and dispersal among
connected patches exhibits the mechanism of generation of
oscillations, i.e., rhythmogenesis, as well as suppression of
oscillations, i.e., amplitude death and oscillation death.  Also, the
transition of homogeneous steady state to inhomogeneous steady state
occurs through a co-dimension-2 bifurcation.  Emphasizing a network of
spatially extended system, the coupled model exposes the collective
behavior of synchrony-stability relationship with various
synchronization occurrences such as in-phase and out-of-phase.
\end{abstract}



\maketitle

\section{Introduction}
For increasing complexity in ecosystems, modeling ecological
consequences in continuously changing environmental conditions is one
of the central concerns of theoretical ecologists. Various key
components like environmental heterogeneity, habitat fragmentation,
habitat loss and seasonal pattern or climatic change profoundly impact
many biological phenomena such as synchrony of oscillating
populations, community structure, diversity-stability and
synchrony-stability relationship \cite{Han98,Han_book,GoHa08}. In
particular, habitat connectivity through species dispersal among
fragmented landscapes play significant role in determining ecosystem
functioning and evolutionary processes
\cite{Na08,DaBr08,Coli12}. Moreover, in spatial ecology, the
connectivity of habitats conserves the ecological system by balancing
the natural conditions \cite{Han99}. As population movement prevents
local species loss from complete extinction in their local habitat, so
it is important to understand the factors which influences the
dispersal effect. Numerous coupled nonlinear systems and coupled
stochastic oscillators associated with dispersal conceptualize various
biological notions starting from oscillations to chaotic behavior
\cite{KeAk08,BaDu15,RaDuBa15}.  Particularly, in the context of
species population dynamics, spatially extended dynamical systems with
environmental heterogeneity play leading role in determining the
community structure and in maintaining the biodiversity in regional
landscapes \cite{Van06,KeMa00,Han_book}.

Much less is still known about the effects of species dispersal in
dynamic habitats of ecosystems. Recently, in physical generic
oscillators, dynamic environment has been considered in a scheme of
indirect coupling to show various synchronization behavior
\cite{Kat08} as well as synchronization of chaotic systems
\cite{ReAm10}, amplitude death and oscillation death
\cite{ReAm11,ReAm12,GhBa14}.  As far as ecological systems are
concerned, in general, habitats are highly dynamic rather than static
\cite{Han99,KeMa00} where species dispersal takes place in order to
maintain the species diversity and persistence \cite{Gu02}.  In search
of preferential food, sometimes species move a long distance for
suitable habitats \cite{MuFe13}.  While moving from one habitat to
another habitat, species get the available food from the environment
for their survival.  Considering the varying environmental conditions
as a dynamic variable, we model the consumer--resource interaction in
a dynamic environment along with the presence of dispersion.  We
consider ecological oscillators represented by the
Rosenzweig--MacArthur model under the simultaneous influence of two
different types of couplings: One is the {\it direct coupling}
describing the dispersal of species between spatially separated
patches and another one is the {\it indirect coupling} describing a
common dynamic environment for the patches which are connected via
dispersal.  As far as ecological environment is concerned, we consider
the logistic growth model replicating a dynamic landscape.  Our main
concern in this paper is to investigate how the interplay of these two
separate types of coupling affects the collective dynamical behavior
of an ecological system.

In general, the concepts of nonlinear dynamics have been widely used
in many studies on complex ecological systems to characterize various
natural processes and its ecological perspectives
\cite{Str_book,Mur_book}.  Emphasizing that, in coupled nonlinear
systems, {\it rhythmogenesis} is an interesting dynamical phenomenon,
such that the interaction through underlying coupling generates the
oscillation from their respective steady states. Generation of rhythms
from steady state is a novel phenomenon in most of the coupled
physical, chemical and biological systems \citep{Das10, Chak15,
  Gla01}. Especially, for regulation and restoration of oscillatory
behaviors in various physiological and neuronal systems,
rhythmogenesis is utilized as an important consequence of coupled
dynamical systems in which the underlying coupling acts as a feedback
factor \citep{Zou15, Loe01, Harr10} and drives the system away from
steady state to oscillations.  In contrast to the generation of
oscillation from steady state, specific coupling involved in the
system can drive coupled oscillators to oscillation quenching states,
such as {\it oscillation death} (OD) and {\it amplitude death}
(AD). In coupled oscillators, oscillation death (OD) is created by
suppression of oscillations with formation of stable {\it
  inhomogeneous steady states} whereas AD is created by the
suppression of oscillations with formation of stable {\it homogeneous
  steady states} \cite{KoVo13,KoVo13a}.  As far as ecological models
are concerned, oscillation quenching mechanisms emphasize the
relationship between dispersal induced stability and dispersal induced
synchrony.  Using mean-field coupling as species dispersal in
spatially extended ecological  model, the relationship between
synchrony and stability induced by dispersal has been shown in
Ref.\cite{BaDu15,RaDuBa15,DuBa15}.

In this paper, through a rigorous bifurcation analysis we show that
environmentally coupled ecological systems exhibit both generation and
suppression of oscillations, which are two diametrically opposite
phenomena.  Using this environmentally coupled model in homogeneous
patches (i.e., identical patches), we show amplitude death,
oscillation death, formation of inhomogeneous limit cycles and
transition of AD to OD through a co-dimension-2 bifurcation.  In
particular, we emphasize that rhythmogenesis can occur along with
different synchronized behavior.  Although, the uncoupled RM model has
simple characteristics, but due to presence of environmental coupling
and dispersal among the fragmented habitats, the system exhibits many
interesting dynamics which can be useful in the context of ecological
systems.  Moreover, we show that the dynamics of the considered system
are qualitatively valid for a network of patches connected by a common
environment and the system shows rhythmogenesis, perfect synchrony,
in-phase synchrony, out-of-phase synchrony and the formation of
multi-cluster state for a network of nearby connected patches.

We organize the paper as: First, the environmentally coupled
ecological model along with the dynamics of uncoupled system as well
as the dynamics of environment are described in Sec.~\ref{S:2}.  Based
on this model, the effect of dispersal and it's dynamics are shown in
Sec.~\ref{S:3} in two different ways. One is based on the assumption
that uncoupled patch is in oscillatory state and second one is started
with steady state in uncoupled patch. Further, we extend to a network
of ring coupled oscillators and robustness of this model is shown in
Sec.~\ref{S:4}.  Finally, we discuss the results from dynamical
systems point of view and also its ecological interpretation in
Sec.~\ref{S:5}.

\section{Mathematical Model}
\label{S:2}

Taking into account the dispersal between two spatially separated
patches in a dynamic environment, we model the consumer--resource
interaction in a single system emphasizing the dynamics of consumer
interaction in three ways. First one is the consumer interaction
within the patch, second is consumers interaction between the
connected patches through dispersal, and finally, the consumer
interaction through a common dynamic environment.  We consider,
dynamics of the consumer ($H$) and the resource ($V$) interaction
within a patchy habitat is given by the Rosenzweig--MacArthur (RM)
model \cite{RoMa63,Mur_book}.  Now, for the interaction between the
isolated patches, we consider dispersal of consumer populations
between spatially isolated patches.  In fact, immigration and
emigration are set by directly coupling the patches with dispersal
rate ($d$).  Finally, while dispersing between patches, the consumer
interacts with environment and gets the available resource from the
environment for their survival during dispersal.  This is potentially
important for long distance dispersal of species.  To describe the
interaction through the environment, we consider another resource
($E$) as a common dynamic environment and uncoupled dynamics of the
environment is given by the logistic growth model.  Further, the
interaction between consumers ($H_{1,2}$) and environment is governed
by a type-II functional response \cite{Ho59,Mur_book}.  The choice of
this particular function is motivated by the fact that it represents a
functional response for many consumers. The shape of the function is
based on the idea that, at low resource densities, consumers spend
most of their time on searching for resources, whereas at high
resource densities, they spend most of their time on resource handling
\cite{Ho59}.  The dynamics of consumer ($H$) and resource ($V$) for
  patch-1 and patch-2 along with the common environment ($E$) are
  modeled as:
\begin{subequations}\label{eq1}
\begin{align}
 \frac{dV_{1,2}}{dt} &= r V_{1,2} \left(1-\frac{V_{1,2}}{K}\right)-
 \frac{\alpha V_{1,2}}{V_{1,2}+B} H_{1,2}\; , \\ \frac{dH_{1,2}}{dt} &=
 H_{1,2} \left(\beta\frac{\alpha V_{1,2}}{V_{1,2}+B}-m\right) +
 d(H_{2,1}-H_{1,2}) \nonumber \\ ~~& ~~~~~~~ ~~~~\;\;\;+~ \gamma
 \frac{ \epsilon E}{E+C} H_{1,2}\;,\\ \frac{dE}{dt} &= r_1 E
 \left(1-\frac{E}{K_1}\right)- \frac{\epsilon E}{E+C}(H_{1}+H_2)\;,
\end{align}
\end{subequations}
where $r$ is the growth rate of the resource ($V$) with $K$ as its
carrying capacity, $\alpha$ and $B$ represent the predation rate of
the consumer ($H$) and half saturation constant respectively. The
predation efficiency and mortality rate of the consumer ($H$) are
respectively given by $\beta$ and $m$.  The environmental resource
($E$) acts as another resource consisting of logistic growth
dynamics. In this case, the growth rate and the carrying capacity of
the environment ($E$) are given by $r_1$ and $K_1$, 
respectively. Further, $\epsilon$ and $\gamma$ represent the predation
rate and the predation efficiency of the consumer ($H$) due to
available resource from the common environment. In predatory level,
$C$ is half saturation constant of the environment ($E$).  The
equations (\ref{eq1}a)-(\ref{eq1}c) emphasize that we use {\it direct
  coupling} between the consumers from neighboring patches via the
dispersal rate $d$ and {\it indirect coupling} through the predation
rate $\epsilon$ of the consumer ($H$).

In all the previous studies \cite{Kat08,ReAm10,ReAm11,ReAm12,GhBa14}
the environment has been considered as an overdamped oscillator having
no intrinsic dynamics other than zero steady state: the environment is
modulated in a linear way in the presence of systems that are coupled
with it. This is an oversimplified view of any environment and thus,
the true essence of (nonlinear) dynamical environment was missing in
those studies. As the environment plays a key role in determining the
dynamics of the coupled system, therefore, it is of fundamental
interest to study the effect of an environment that is represented
with some {\it realistic} model. Here, not only we take a realistic
environment (logistic growth model) but, apart from that, unlike the
previous studies \cite{ReAm10,ReAm11,ReAm12,GhBa14} we also consider
the interaction among the environment and systems to be nonlinear.

Interestingly, it can be shown that the model of environment
considered in \cite{ReAm10,ReAm11,ReAm12,GhBa14} viz. $\frac{dE}{dt} =
-\mathcal{K}E$, where $\mathcal{K}$ is the damping parameter, is a
special case of the general model given in Eq.~(\ref{eq1}c).  When
$\epsilon=0$, from Eq.~(\ref{eq1}c) the per capita rate of change of
environment is
$\frac{1}{E}\frac{dE}{dt}=\frac{f(E)}{E}=r_1(1-\frac{E}{K_1})$ and
hence declines by the quantity $-r_1/K_1$ for each individual added in
the population.  Let us consider that we perturb $E^*$ by a small
quantity $s$, so the perturbed density is $E^*+s$. Now $\frac{dE}{dt}
= \frac{d(E^*+s)}{dt} =\frac{dE^*}{dt}+\frac{ds}{dt} =
\frac{ds}{dt}=f(E^*+s)$ as $\frac{dE^*}{dt}=0$.  Expanding the
function $f(E^*+s)$ about $E^*$ and neglecting the higher order terms
we get $f(E^*+s)= f(E^*)+s\frac{df}{dE}|_{E^*} $.  As $f(E^*)=0$ and
$\frac{df}{dE}|_{E^*}=-r_1$, so the approximate form of
$\frac{dE}{dt}$ a small distance away from the equilibrium point
$E^*=K_1$ is $\frac{ds}{dt}=-r_1s$, where $r_1>0$.  Hence, if the
steady state of the environment is perturbed by a small amount by some
external factors it will actually behave like an overdamped
oscillator.

Thus, the environment we have considered here resembles the overdamped
model for a smaller perturbation, but additionally, it also
accommodates the effect of larger perturbation exerted by the
systems. Therefore, the environment considered here, indeed,
represents a general model of environment. In the following we will
emphasize on the results that come due to this nonlinear environment,
and which were not observed for the previously considered simplified
overdamped model of environment.

\section{Stability Analysis}

The uncoupled dynamics of the RM model (i.e.,
Eqs.~(\ref{eq1}a)-(\ref{eq1}b) with $d=0$ and $\epsilon=0$) are
discussed in \cite{RoMa63,BaDu15}.  Here we briefly discuss it for the
sake of clarity. The RM model has two trivial equilibrium points
(i.e., $(V^*,H^*)=(0,0)$ and $(V^*,H^*)=(K,0)$) and one non-trivial
equilibrium point (i.e.,
$(V^*,H^*)=\left(\frac{mB}{\alpha\beta-m},\frac{B r \beta(K(\alpha
  \beta-m)-Bm)}{K(\alpha\beta-m)^2}\right)$). This non-trivial
equilibrium point is stable if $\frac{B}{K} <
\frac{\alpha\beta-m}{\alpha\beta+m}$ for certain parameter values of
$K$. However, for a range of $K$ value, the non-trivial equilibrium
point changes it's stability and raises to stable limit cycle. Hence,
the RM model is in either steady state or in oscillation state
depending upon different parametric values.

Now, let us investigate the stability of the coupled system given by
Eq.~(\ref{eq1}).  The trivial equilibrium points of the environmentally
coupled system (\ref{eq1}) are given by $(0, 0, 0, 0, 0)$, $(K, 0, 0,
0, 0)$, $(0, 0, K, 0, 0)$, $(0, 0, 0, 0, K_1)$, $(K, 0, 0, 0, K_1)$,
$(0, 0, K, 0, K_1)$, $(K, 0, K, 0, K_1)$, $(V^{o}, H^o, V^o, H^o, 0)$
and a non-trivial equilibrium point is given by ($V^*, H^*, V^*, H^*,
E^*$). As the expression of the non-trivial equilibrium point is quite
cumbersome, here we don't give the expression. However, the
expressions for $V^o$ and $H^o$ are given by:
\begin{eqnarray*}
V^o & =  & \frac{Bm}{\alpha \beta - m},~~ \mbox{and}  \\ H^o & = &
-\frac{B r \beta (B m + K m -K \alpha \beta)}{K(\alpha \beta-m)^2}\;.
\end{eqnarray*}

\begin{figure*}
\includegraphics[width=0.99\textwidth,bb=-200 194 843 667]{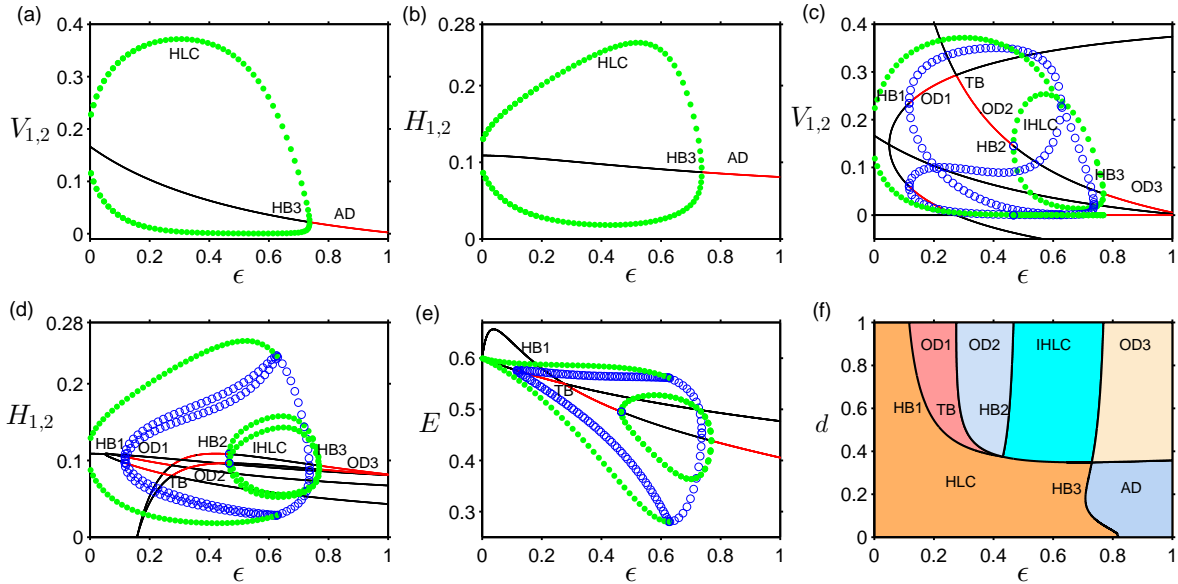}
\caption{(Color online) Oscillation and amplitude death: (a)--(b) One
  parameter bifurcation diagram for varying predation rate
  ($\epsilon$) with lower value of dispersal rate ($d=0.2$).  (c)--(e)
  One parameter bifurcation diagram for varying predation rate
  ($\epsilon$) with higher dispersal rate ($d=1$). Red and black
  curves represent stable and unstable steady states respectively,
  whereas green and blue circles represent stable and unstable limit
  cycles respectively.  Here OD, AD, HB, TB, IHLC and HLC represent
  oscillation death, amplitude death, Hopf bifurcation, transcritical
  bifurcation, inhomogeneous limit cycle and homogeneous limit cycle
  respectively.  (f) Two parameter bifurcation diagram in $\epsilon-d$
  plane.  Other parameters are $r=0.5$, $K=0.5$, $\alpha=1$, $B=0.16$,
  $\beta=0.49$, $m=0.25$, $\gamma=0.6$, $r_1=0.5$, $K_1=0.6$ and
  $C=0.6$.}
\label{f:syn}
\end{figure*}

The Jacobian matrix ($\mathcal{J}$) at the non-trivial equilibrium
point $(V^*, H^*, V^*, H^*, E^*)$ is given by:
$$\displaystyle \mathcal{J}|_{(V^*, H^*, V^*, H^*, E^*)}=
  \begin{bmatrix}
    j_{11}& j_{12} &:& 0 &0 & : & 0\ \\ j_{21} & j_{22} &:& 0&j_{24} &
    : & j_{25} \\ \cdots & \cdots & & \cdots & \cdots& & \cdots
    \\ 0&0&:& j_{11} & j_{12} & : & 0\\ 0& j_{24} &:& j_{21} &j_{22} &
    : & j_{25} \\ \cdots & \cdots & & \cdots & \cdots& & \cdots \\ 0 &
    j_{52} &:& 0 &j_{52} & : & j_{55}
 \end{bmatrix},$$
where
\begin{eqnarray*}
 j_{11}&=& r - \frac{2 r V^*}{K} - \frac{B \alpha H^*}{(B + V^*)^2},
 ~~ j_{12}=-\frac{ \alpha V^*}{(B + V^*)} \\ j_{21} &=&\frac{B \alpha
   \beta H^*}{(B + V^*)^2},~~j_{22} = -d - m + \frac{\alpha \beta
   V^*}{(B + V^*)} + \frac{\gamma \epsilon E^* }{(C + E^*)}
 \\ j_{24}&=&d,~~j_{25}= \frac{c \gamma \epsilon H^*}{(C +
   E^*)^2},~~j_{52}=\frac{\epsilon E^*}{(C + E^*)}\;\; \mbox{and}\\j_{55}&=&
 r_1-\frac{2r_1E^*}{K_1} + \frac{2C\epsilon H^*}{C+E^*}\;.
\end{eqnarray*}
In the Jacobian $\mathcal{J}$, we have block matrices which simplifies
the method of finding eigenvalues and the corresponding
eigenvalues are given by the expression:
\begin{eqnarray*}
 \lambda_{1,2}&=& \frac{1}{2} \left(j_{11}+j_{22}-j_{24} \mp
 \sqrt{4j_{12}j_{21}+(j_{11}-j_{22}+j_{24})^2} \right)
\end{eqnarray*} 
and $\lambda_{3,4,5}$ can be found out by the roots of the polynomial
$x^3+ax^2+bx+c=0$, where
\begin{eqnarray*}
 a&=& -(j_{11}+j_{22}+j_{24}+j_{55}),
 \\ b&=&-j_{12}j_{21}-2j_{25}j_{52}\\&&+(j_{22}+j_{24})j_{55}+j_{11}(j_{22}+j_{24}+j_{55}),\; \mbox{and}\\ c&=&2j_{11}j_{25}j_{52}+j_{12}j_{21}j_{55}-j_{11}j_{55}(j_{22}+j_{25}).
\end{eqnarray*}
From these eigenvalues, stability of the equilibrium point for
different parameter values can be found.  However, it has to be noted
that due to the nonlinear nature of the coefficients of Jacobian, in
most of the cases it is very difficult to predict the exact
bifurcation points using eigenvalue analysis. Thus, to locate the
bifurcation points and curves, we resort to the continuation package
$\mbox{XPPAUT}$ \cite{xpp}. The collective behavior of the coupled
systems for different coupling and system parameters are described in
the next section.

\section{Results}
\label{S:3}

Emphasizing that, without coupling the RM model has either steady
state or oscillatory state depending on the values of parameters, we
qualitatively describe the coupled dynamics of system (\ref{eq1}) in
two distinct conditions: First, we consider the uncoupled patches are
in oscillatory condition and examine how coupling affects the
oscillation suppression states.  Whereas, in the second case we
consider the uncoupled patches are in equilibrium condition and
investigate how coupling changes this quiescent state in order to
establish any possible rhythmogenesis. In the numerical bifurcation
analysis \cite{xpp} we use the fourth-order Runge-Kutta algorithm with
step size 0.001.

\subsection{Uncoupled patches are in oscillatory state}
We start with the precondition that in the absence of dispersal (i.e.,
$d=0$) between the patch-1 and patch-2 and also the in the absence of
coupling with environment (i.e., $\epsilon=0$), the resource ($V$) and
the consumer ($H$) show oscillations.  Then, we study the simultaneous
effects of dispersal and environmental coupling.

\subsubsection{Amplitude death in identical oscillators}

For lower dispersion rate determined by the parameter $d$, we observe
that with increasing predation rate $\epsilon$, AD appears from stable
limit cycle through an inverse supercritical Hopf bifurcation
(HB3). Figures~\ref{f:syn}(a) and \ref{f:syn}(b) show the AD state
beyond $\epsilon_{HB3}=0.7367$ for the resource ($V_{1,2}$) and the
consumer ($H_{1,2}$), respectively for $d=0.2$. Here in both patch-1
and patch-2, populations are suppressed to homogeneous steady states.

In Ref.~\cite{GhBa14} it has been shown that an environment modeled by
an overdamped oscillator along with the linear modulation from the
system can induce AD in the presence of dispersion in generic
oscillators. But, the region of AD in the $d-\epsilon$ space was shown
to be very narrow. In contrast to that, in our present case once AD
occurs at $\epsilon_{HB3}$ the system stuck to that state even for
higher values of $\epsilon$. This contrast in results may be
attributed to the fact that, unlike Ref.~\cite{GhBa14}, here we
consider a nonlinear environment and nonlinear coupling. For higher
$d$, symmetry breaking is resulted, which is discussed below.

\begin{figure}
\includegraphics[width=0.41\textwidth]{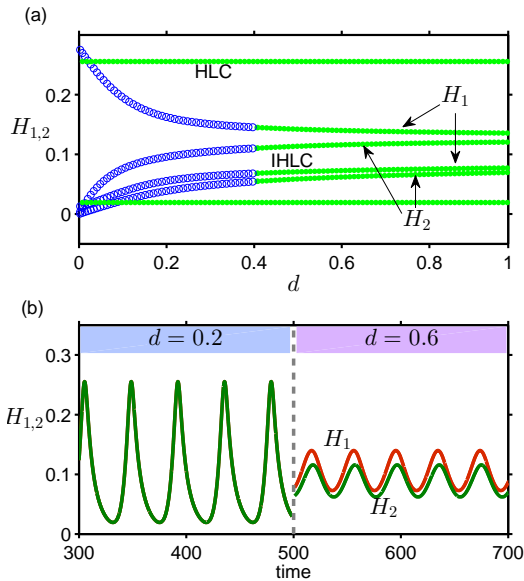}
\caption{(Color online) Transition from HLC to IHLC: (a) One
    parameter bifurcation diagram of $H_{1,2}$ with varying dispersal
    rate $d$.  (b) Time series of $H_{1,2}$ which exhibits the direct
    transition from HLC to IHLC with change in $d$. We consider
    $\epsilon=0.5$ and all the other parameters are same as in
    Fig.~\ref{f:syn}.}
\label{f:hlc}
\end{figure}

\subsubsection{Oscillation death for higher dispersal rate ($d$)}

Beyond a certain dispersal rate $d$, for increasing predation rate
($\epsilon$), IHSS is created at $\epsilon_{PB}=0.05023$: this IHSS
gets stable and gives rise to oscillation death (OD1) at
$\epsilon_{HB1}=0.117$.  Also, OD2 is created from OD1 through a
transcritical bifurcation (TB) at $\epsilon_{TB}=0.2752$.  It is
noteworthy that, OD1 and OD2 are accompanied by a stable limit cycle,
but OD3 that is created through supercritical Hopf bifurcation (HB3)
for a higher dispersal value does not share the phase space with any
oscillatory state.  Moreover, for $m=0.25$, the dynamics of the
environmental resource ($E$) is shown in Fig.~\ref{f:syn}(e).
Although, uncoupled dynamic environment ($E$) is in steady state, it
shows the synchronized oscillating behavior when its coupled with the
oscillating patches.

In Fig.~\ref{f:syn}(c), OD1 shows inhomogeneous steady states where
$V_1,V_2,H_1, H_2$ and $E$ have non-zero density. But, in the OD2
state, the density of the resource ($V_i$) in one patch is almost zero
and another is non-zero.  Surprisingly, in this situation, the
consumers ($H_1$ and $H_2$) from both patches have non-zero density
both in OD1 and OD2 states.  Even though the resource ($V_i$) in one
patch of OD2 is almost zero, but the consumer ($H_i$) from that patch
survives because of the available resource in the environment.
Indeed, in the absence of resource ($V_i$), the survival of consumer
($H_i$) is completely supported by the environmental resource ($E$)
only.

The importance of direct coupling and environmental coupling and it's
dynamics are shown for broader range of parameters using two parameter
bifurcation diagram in $\epsilon - d$ space.  Figure~\ref{f:syn}(f)
shows the parameter region where the oscillation quenching takes place
in directly and indirectly coupled RM model.  Each shaded region in
Fig.~\ref{f:syn}(f) determines the dynamics of the coupled system for
distinct values of $\epsilon$ and $d$. Moreover, when the dispersal
rate ($d$) is more than $d_{PB}=0.3519$ (approximately), OD1, OD2,
IHLC and OD3 occur in the system, whereas only amplitude death occurs
for low dispersal rate.

\begin{figure}
\includegraphics[width=0.465\textwidth]{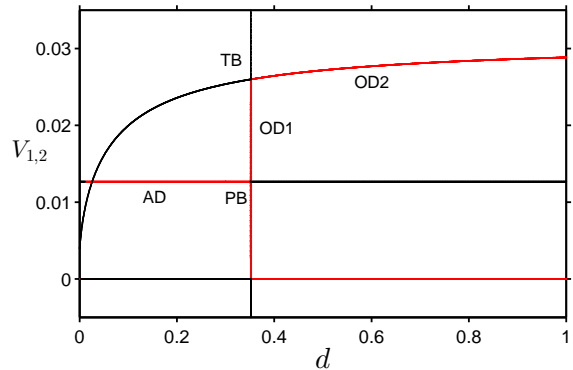}
\caption{(Color online) AD--OD transition: One parameter bifurcation
  diagram for varying $d$. Here PB represents Pitchfork
  bifurcation. Red and black curves represent stable and unstable
  steady states respectively. Other fixed parameters: $\epsilon=0.85$,
  $\alpha=1$, $r=0.5$, $K=0.5$, $B=0.16$, $\beta=0.49$, $r_1=0.5$,
  $K_1=0.6$, $C=0.6$, $\gamma=0.6$ and $m=0.25$ }
\label{f:adod}
\end{figure}

\begin{figure*}
\includegraphics[width=1.02\textwidth,bb= -200 271 843 560]{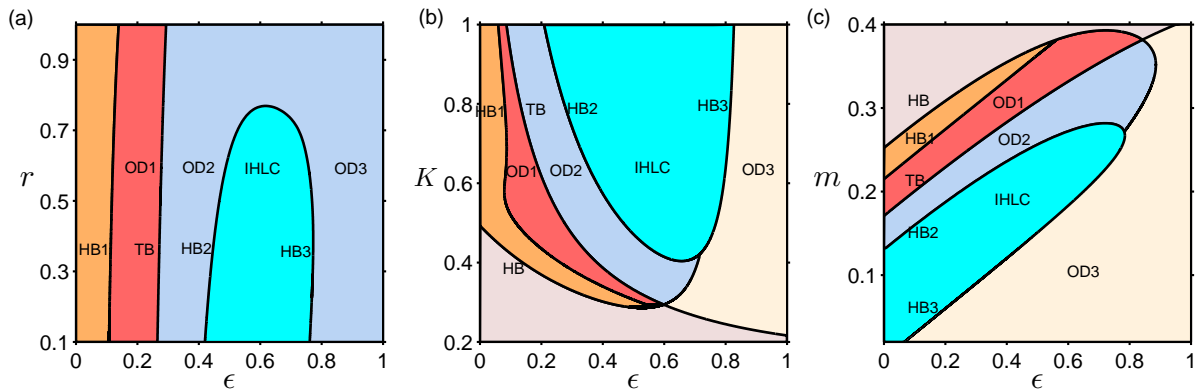}
\caption{(Color online) Local system parameters with environmental
  coupling parameters: (a) $\epsilon - d$ space as a two parameter
  bifurcation diagram, (b) $\epsilon - K$ space, (c) $\epsilon -
  m$ space.  Here fixed parameters are $d=1$, $\alpha=1$, $B=0.16$,
  $\beta=0.49$, $r_1=0.5$, $K_1=0.6$, $C=0.6$ and $\gamma=0.6$. In
  each diagram, other fixed parameters are $r=0.5$, $K=0.5$ and
  $m=0.25$.}
\label{f:local}
\end{figure*}

\subsubsection{Transition from HLC to IHLC} 

An interesting finding from Fig.~\ref{f:syn}(f) is the transition from
homogeneous limit cycle (HLC) to inhomogeneous limit cycles (IHLC)
with the variation of dispersal rate $d$ for a certain range of
$\epsilon$. The genesis of this transition can be understood more
clearly from the one dimensional bifurcation diagram with $d$ for an
exemplary value $\epsilon=0.5$. From Fig.~\ref{f:hlc}(a) it is seen
that for $\epsilon=0.5$ there exist HLCs even at $d=0$; additionally
they are accompanied by unstable limit cycles [shown in open (blue)
  circles)]. As $d$ is increased, at $d_{HB}$ the unstable limit
cycles get stable to give birth to IHLCs through subcritical Hopf
bifurcation, but the original HLCs remain there beyond that
point. Thus, beyond $d_{HB}$ we can get a direct transition from HLC
to IHLC for an appropriate choice of initial
conditions. Figure~\ref{f:hlc}(b) shows this transition in time series
of HLC (at $d=0.2$) and IHLC ($d=0.6$) [we take $\epsilon=0.5$].

In the $\epsilon$ space (for a fixed $d$), it can be seen from
Figs.~\ref{f:syn}(c) and \ref{f:syn}(d), that inhomogeneous limit
cycles (IHLC) are formed at $\epsilon_{HB2}=0.467$ through a
supercritical Hopf bifurcation (HB2) without considering any mismatch
in species local dynamics. Further this IHLC are suppressed to
inhomogeneous steady states which give raise to OD3.  

\begin{figure*}
\includegraphics[width=0.97\textwidth,bb=-180 174 844 667]{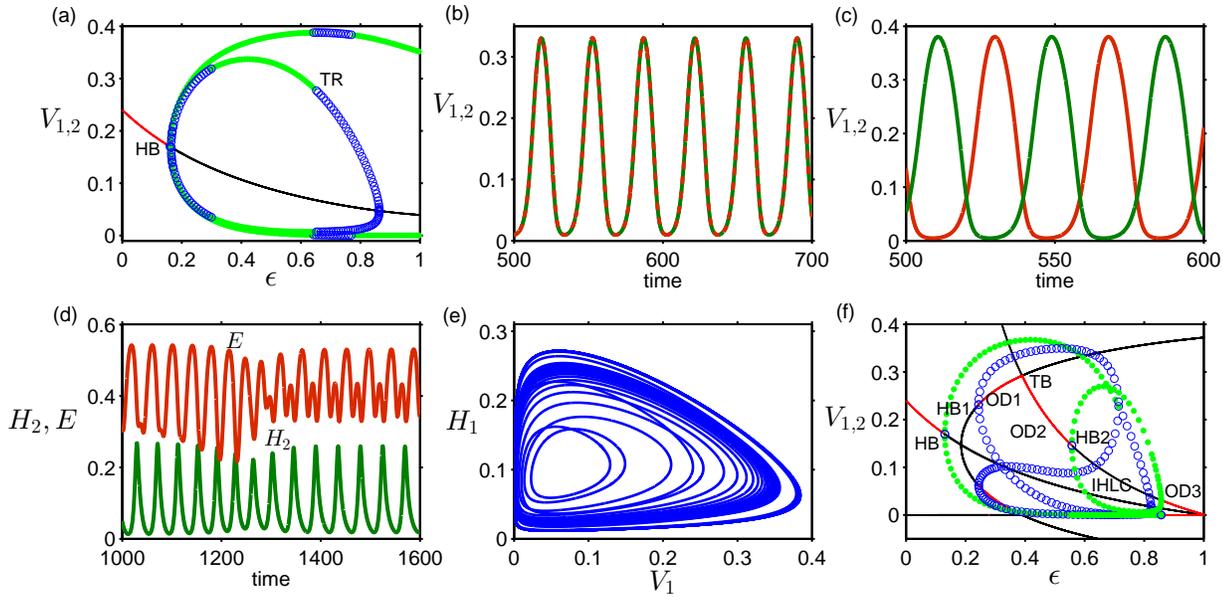}
\caption{(Color online) (a) One parameter bifurcation diagram for
  varying $\epsilon$ with $d=0$. TR represents Torus
  bifurcation. Other fixed parameters are $\alpha=1$, $r=0.5$,
  $K=0.5$, $B=0.16$, $\beta=0.5$, $r_1=0.5$, $K_1=0.6$, $C=0.5$,
  $\gamma=0.5$ and $m=0.3$.  (b) Perfect synchrony for $\epsilon=0.5$
  with the initial condition $(V_1,H_1,V_2,H_2,E)=(0.28, 0.051, 0.24,
  0.052, 0.55)$.  (c) Out-of-phase synchrony for $\epsilon=0.5$ with a
  different initial condition $(V_1,H_1,V_2,H_2,E)=(0.297, 0.145,
  0.016, 0.0308, 0.5)$.  (d) Time series of chaotic behavior for
  $\epsilon=0.76$ (after the Torus bifurcation (TR)).  (e) Phase space
  of the chaotic time series. In (b)--(e), parameters are same as in
  (a).  (f) One parameter bifurcation diagram for varying $\epsilon$
  with fixed $d=1$. Here other parameters are $\alpha=1$, $r=0.5$,
  $K=0.5$, $B=0.16$, $\beta=0.5$, $r_1=0.5$, $K_1=1$, $C=0.5$,
  $\gamma=0.5$ and $m=0.3$.}
\label{f:sty}
\end{figure*}

\subsubsection{AD to OD transition}

From Fig.~\ref{f:syn}(f), it is interesting to note that for a higher
$\epsilon$ value (organized by HB3 curve) the system shows either AD
or OD depending upon the dispersal rate ($d$). Thus, if we fix the
value of $\epsilon$ in this oscillation suppressed region and vary $d$
then a continuous direct transition from AD to OD is observed. But,
unlike other AD-OD transition, here the dynamics is governed by the
co-dimension two bifurcation.

This is shown more clearly in Fig.~\ref{f:adod}, which shows the
direct transition of AD to OD with the symmetry breaking of steady
state through pitchfork bifurcation (PB). For low dispersal rate,
system (\ref{eq1}) shows AD whereas for increasing dispersal rate, OD1
occurs at $d_{PB}=0.3519$. Further an increase in dispersal ($d$), OD2
is created through transcritical bifurcation (TB) at $d_{TB}=0.3519$
which is same as the pitchfork bifurcation point at $d_{PB}=0.3519$.
This shows that $d=0.3519$ is a point of co-dimension 2 bifurcation.
It is important to note that consumers from patch-1 and patch-2
($H_{1,2}$) and also the resource $(V_{1,2})$ are in non-zero density
in OD1 state whereas the resource ($V_{1,2}$) density in OD2 is in low
value.

\subsubsection{Effect of local dynamics}

Relative to the coupling parameters (i.e.,
$d$ and $\epsilon$), other parameters representing behavioral,
morphological and life history traits (characteristics) such as growth
rate ($r$), carrying capacity ($K$) and mortality rate ($m$) also
significantly contribute in the dynamics of spatial ecosystems.  The
resource ($V$) from both patch-$1$ and patch-$2$ doesn't involve in
both direct and environmental coupling.  However, it indirectly plays
the role in determining the oscillation death and amplitude death.
Here spatial variation is taken into account, we show the two
parameter bifurcation diagram for varying the local parameters (i.e.,
$r$, $K$ and $m$) along with varying predation rate ($\epsilon$).

In Fig.~\ref{f:local}, two parameter bifurcation diagrams in $\epsilon
- r$ plane, $\epsilon - K$ plane and $\epsilon - m$ plane are shown
with different color shaded regions representing distinct dynamics of
the system (\ref{eq1}) for different parameters.
Figures~\ref{f:syn}(a)-(e) depict one particular case of
Fig.~\ref{f:local} with distinct change in each parameter.  Here
change in growth rate ($r$) does not affect much in the OD1 region
(shown in Fig.~\ref{f:local}(a)) but changes the OD2 region by
shrinking inhomogeneous limit cycles. In fact, for higher growth rate
($r$), the inhomogeneous limit cycles vanish and further OD2 and OD3
coincides together and thus forms only one OD.  Same conclusion holds
for decreasing the carrying capacity ($K$).  For low value of the
carrying capacity, no oscillation occurs (shown in
Fig.~\ref{f:local}(b)). As far as local dynamics of the patch is
concerned, mortality rate ($m$) is an important factor in
heterogeneous fragmented landscapes. We check the dynamics for varying
mortality rate and predation rate. For increasing mortality state, OD
region and inhomogeneous limit cycles disappear (shown in
Fig.~\ref{f:local}(c)).  At low mortality rate, IHLC and OD occur
whereas for higher mortality rate, we have only homogeneous limit
cycles along with oscillation death through HB3.  Importantly, both
environmental coupling parameters and local dynamical parameters
determine the dynamics of the considered system.

\subsection{Uncoupled patches are in equilibrium state}
All the results shown in the previous section are based on the initial
assumption that the dynamics of uncoupled system is in oscillatory
state.  Now we explore the effect of direct and indirect coupling when
the uncoupled systems are in equilibrium state.  We set the individual
uncoupled systems in steady state by choosing proper value of
mortality rate ($m$) of the consumer ($H$).

\subsubsection{Rhythmogenesis and oscillation death}
In this section, we set that each patch is in equilibrium state by
choosing the mortality rate of the consumer as $m=0.3$. We check the
dynamics of the system (\ref{eq1}) in two different cases. One is
absence of direct coupling and another is considered with direct
coupling along with coupling through the environment.\\

\begin{figure*}
\includegraphics[width=0.84\textwidth]{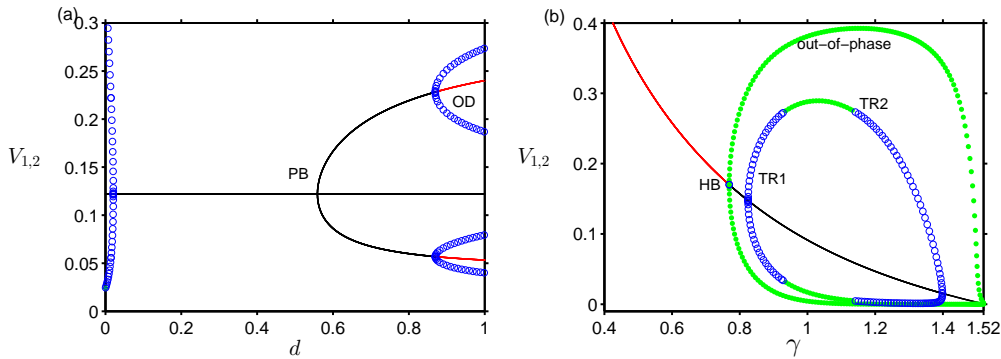}
\caption{(Color online) Direct coupling: (a) one parameter
  bifurcation diagram for varying $d$ with the fixed parameters
  $\epsilon=0.1636$, $\alpha=1$, $r=0.5$, $K=0.5$, $B=0.16$,
  $\beta=0.5$, $r_1=0.5$, $K_1=0.6$, $C=0.5$, $\gamma=0.5$ and
  $m=0.27$. (b) One parameter bifurcation diagram for varying
  predation efficiency $\gamma$ for fixed parameters $d=0$,
  $\alpha=0.6$, $r=0.5$, $K=0.5$, $B=0.16$, $\beta=0.5$, $r_1=0.5$,
  $K_1=0.7$, $C=0.3$, $\epsilon=0.3$ and $m=0.3$.}
\label{f:D}
\end{figure*}

{\bf \it Case - I : Absence of direct coupling}\\

In the absence of direct coupling (i.e., $d=0$), we show the dynamics
of system (\ref{eq1}) using one parameter bifurcation diagram for
varying $\epsilon$ with resource density in
Fig.~\ref{f:sty}(a). Interestingly, oscillations are created at
$\epsilon_{HB}=0.1605$ in the presence of environmental coupling alone
through supercritical Hopf bifurcation.  The generation of oscillation
through coupling is termed as rhythmogenesis in the literature
\citep{Das10, Chak15}.  Hence, the presence of environmental coupling
drives the steady state to oscillatory state and creates
rhythmogenesis with finite period of oscillation.  Moreover, torus
bifurcation is created here with chaotic dynamics (shown in
Fig.~\ref{f:sty}(a)).  The torus bifurcation is associated with the
Neimark-Sacker bifurcation of the Poincar\'e map of a limit cycle in
ordinary differential equation. Sometimes torus bifurcation is also
referred as secondary Hopf bifurcation and occurs at the double-Hopf
bifurcation of the equilibrium point in continuous dynamical system.
In Fig.~\ref{f:sty}(a), torus bifurcation occurs at
$\epsilon_{TR}=0.652$ with a pair of eigenvalues $0.6907\pm0.7232i$
which has unit modulus.  We check back and forth of this Torus
bifurcation for different $\epsilon$ value. When $\epsilon=0.6518$, a
pair of Floquet multipliers $0.6905\pm0.7224i$ exists with absolute
value $0.9993$ whereas when $\epsilon=0.6528$, a pair of Floquet
multiplier $0.6917\pm0.7267i$ exists with absolute value $1.003$.
Further, corresponding trajectories and phase space of chaotic
dynamics are shown in Figs.~\ref{f:sty}(d) and \ref{f:sty}(e).

\begin{figure*}
\includegraphics[width=0.84\textwidth]{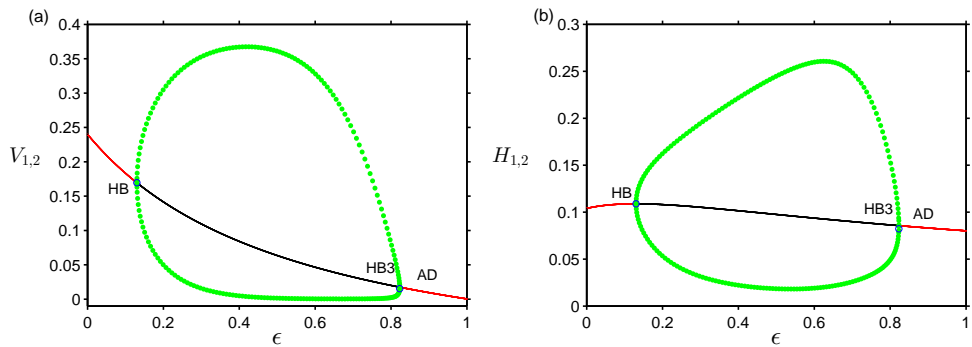}
\caption{(Color online) Amplitude Death :(a) -(b) One parameter
  bifurcation diagram for varying $\epsilon$ for the fixed parameters
  $\alpha=1$, $r=0.5$, $K=0.5$, $B=0.16$, $\beta=0.5$, $r_1=0.5$,
  $K_1=1$, $C=0.5$ and $m=0.3$. Other fixed parameters are
  $\gamma=0.5$ and low value in direct coupling strength $d=0.3$.}
\label{f:AD}
\end{figure*}

As there is no dispersal (i.e., no direct coupling) between the
patches, only environmental coupling plays role in determining the
dynamics here. In Fig.~\ref{f:sty}(a), it is shown that two limit
cycles occur together with a torus bifurcation.  The upper one is in
out-of-phase synchrony, whereas lower one shows the perfect synchrony
between the patches. In fact, in lower one, torus bifurcation occurs.
From this two limit cycles, depending on the initial conditions, we
get either perfect synchrony or out-of-phase synchrony.
Figures~\ref{f:sty}(b) and \ref{f:sty}(c) show time series of perfect
and out-of-phase synchrony of the resource ($V_{1,2}$) for fixed
$\epsilon=0.5$, but with different initial conditions.  For increasing
predation rate ($\epsilon$), the system shows the chaotic behavior and
perfect synchrony.  At $\epsilon=0.76$, in Figs.~\ref{f:sty}(d) and
\ref{f:sty}(e) time series and phase space are shown respectively.
  Hence, perfect synchrony, out-of-phase synchrony and chaotic dynamics
  occur with presence of indirect coupling.\\

{\it Case - II : Presence of direct and indirect coupling}\\

With the presence of direct coupling (i.e., $d \neq 0$), occurrence of
oscillation death is shown in Fig.~\ref{f:sty}(f) for increasing the
predation rate ($\epsilon$).  For this case also, rhythmogenesis
occurs through the variations in the coupling parameter $\epsilon$.
In fact, coupling drives the steady state to oscillatory state which
further quenched to oscillation death as mentioned in the earlier
section.  First oscillation are generated through supercritical Hopf
bifurcation at $\epsilon_{HB}=0.1301$ and then OD1 is created at
$\epsilon_{HB1}=0.2438$ by symmetry breaking of steady state through
pitchfork bifurcation. Further, OD2 is created through transcritical
bifurcation at $\epsilon_{TB}=0.3899$.  Moreover, inhomogeneous limit
cycles arise at $\epsilon_{HB2}=0.557$ which is further suppressed to
OD3 at $\epsilon=0.857$ [shown in Fig.~\ref{f:sty}(f)].

For fixed direct coupling parameter $(d)$, we have shown the various
dynamics of system (\ref{eq1}) with variations in environmental
coupling parameters. In contrast, for fixed environmental coupling
parameter, we find the dynamics for varying direct coupling parameter.
We show the one parameter bifurcation diagram for varying the coupling
strength ($d$) in Fig.~\ref{f:D}(a).  Important to note that here we
always start with steady state in uncoupled patches. OD is created at
$d_{OD}=0.8698$ by symmetry breaking of steady state through pitchfork
bifurcation (PB).

\begin{figure*}
\includegraphics[width=0.9\textwidth,bb=-190 255 840 586]{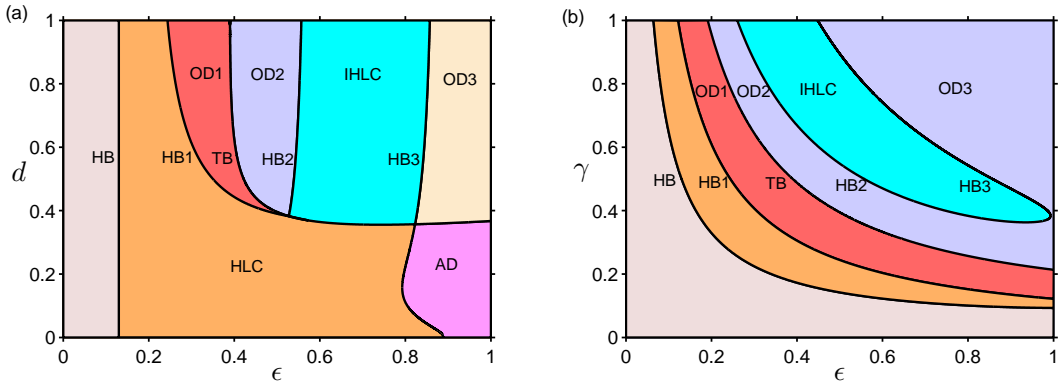}
\caption{(Color online) Two parameter bifurcation diagram: (a)
  $\epsilon - d$ space for fixed $\gamma=0.5$, (b) $\epsilon - \gamma$
  space for fixed $d=1$. Other fixed parameters are $\alpha=1$, $r=0.5$,
  $K=0.5$, $B=0.16$, $\beta=0.5$, $r_1=0.5$, $K_1=1$, $C=0.5$ and
  $m=0.3$.}
\label{f:EG}
\end{figure*}

\subsubsection{Perfect and out-of-phase synchrony}

As we use environmental coupling to connect the indirect interaction
of the consumers between the patches via a common environment,
parameters such as predation rate ($\epsilon$) and predation
efficiency ($\gamma$) in the environmental coupling determines the
intrinsic dynamics of the system.  So far, for changing predation rate
($\epsilon$), we determine the dynamics of this system.  However, for
increasing predation efficiency ($\gamma$), we show other interesting
dynamics using the one parameter bifurcation diagram in
Fig.~\ref{f:D}(b).  As mentioned earlier in Fig.~\ref{f:sty}, here
oscillations are created through coupling and further we have
out-of-phase, in-phase synchrony and perfect synchrony among the
patches.  Also, we have torus bifurcation where the stable limit cycles
transitioned to unstable limit cycles.  For a particular $\gamma$
value, we have perfect and out-of-phase synchrony which occur
completely depending on the initial conditions.

\subsubsection{Rhythmogenesis and amplitude Death (AD)}

As we start with fixed steady state in the uncoupled patch, in all the
cases for an increase in predation rate, oscillations are created and
further suppressed to inhomogeneous steady states. In another
parametric set up and for low value of direct coupling strength ($d$),
we have rhythmogenesis (homogeneous limit cycles) at
$\epsilon_{HB}=0.1301$ and suppressed steady states at
$\epsilon=0.823$ occur for increasing predation date ($\epsilon$). In
this case, one parameter bifurcation diagrams for both consumer and
resource are given in Figs.~\ref{f:AD}(a) and \ref{f:AD}(b),
respectively.

The robustness of the creation of oscillation and oscillation death is
identified by $\epsilon - d$ space and $\epsilon - \gamma$ space in
Figs.~\ref{f:EG}(a) and \ref{f:EG}(b) using two parameter bifurcation
diagram. For low to high dispersal rate ($d$), rhythmogenesis always
occurs at $\epsilon_{HB}=0.1301$ which is a limit point to oscillation
creation. For low dispersal rate, rhythmogenesis and AD take place
whereas for higher coupling strength, both rhythmogenesis, oscillation
death and IHLC occur (shown in
Fig.~\ref{f:EG}(a)).  Also, for increasing predation efficiency in the
environmental coupling, we have death state and inhomogeneous limit
cycles.  However, for higher predation rate and predation efficiency,
only oscillation death (OD3) occurs (shown in Fig.~\ref{f:EG}(b)).

\begin{figure*}
\includegraphics[width=0.94\textwidth,bb=-158 174 743 667]{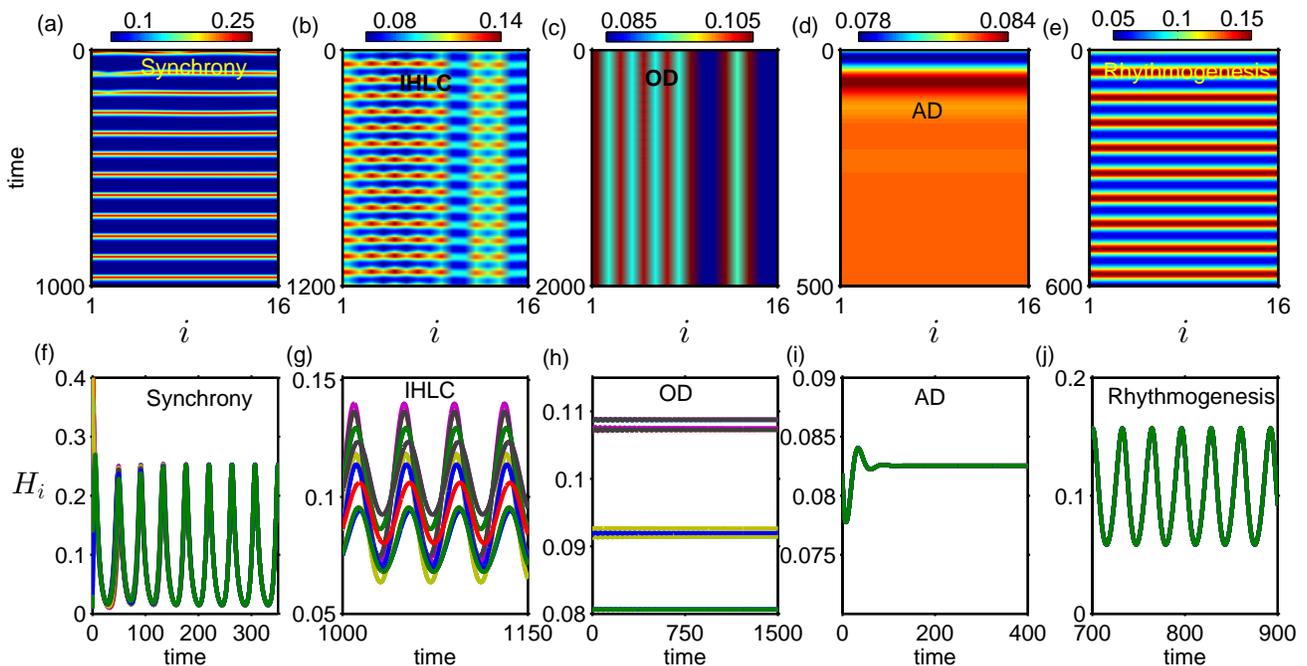}
\caption{(Color online) In a network of 16 patches, spatial dynamics
  and time series: (a,~f) for synchronized oscillations for
  $\epsilon=0.15$, $d=0.35$ and $m=0.2$, (b,~g) synchronized
  oscillations with inhomogeneous limit cycles for $\epsilon=0.2369$,
  $d=0.35$ and $m=0.2$, (c,~h) oscillation death for
  $\epsilon=0.2058$, $d=0.35$ and $m=0.2$, (d,~i) amplitude death for
  $\epsilon=0.55$, $d=0.1$ and $m=0.2$, and (e,~j) rhythmogenesis for
  $\epsilon=0.15$, $d=0.3$ and $m=0.3$.  Other fixed parameters
  are $\alpha=1$, $r=0.5$, $K=0.5$, $B=0.16$, $\beta=0.5$, $r_1=0.8$,
  $K_1=2$, $C=1$ and $\gamma=0.6$ . }
\label{f:NS}
\end{figure*}

\section{Network Structure}
\label{S:4}

We extend the system (\ref{eq1}) to a network consisting $N$ number of
patches coupled by a common dynamic environment ($E$) and also
dispersal takes place among the connected patches. Using a ring type
coupled network (i.e., we consider the periodic boundary condition),
we set that each patch is connected to its nearest neighbor. So
the dynamics of the consumer ($H$) and the resource ($V$) in the
$i^{th}$ patch along with common dynamic environment ($E$) are given
by:
\begin{subequations}\label{eq2}
\begin{align}
 \frac{dV_{i}}{dt} &= r V_{i} \left(1-\frac{V_{i}}{K}\right)-
 \frac{\alpha V_{i}}{V_{i}+B} H_{i} , \\ \frac{dH_{i}}{dt} &=
 H_{i} \left(\beta\frac{\alpha V_{i}}{V_{i}+B}-m\right) +
 d(H_{i+1}-2H_i+H_{i-1})\nonumber \\
& \;\;~~+ \gamma \frac{ \epsilon E}{E+C}
 H_{i},~~~~~~~ i=1,2,\hdots, N,\\ \frac{dE}{dt} &= r_1 E
 \left(1-\frac{E}{K_1}\right)- \sum_{i=1}^{N}\frac{\epsilon E}{E+C}
 H_{i}\; .
\end{align}
\end{subequations}
Since many patches are connected in network with a common environment,
we consider a high carrying capacity ($K_1$) of the environment so
that it can support many connected patches.

We check the robustness of our results at different levels of
dispersal rate for a network consisting 16 patches and a common
dynamic environment.  First for low values of $\epsilon$,
Fig.~\ref{f:NS}(a) shows the synchronized oscillations of 16 patches
when dispersal rate $d=0.35$.  Time series of perfectly synchronized
homogeneous limit cycles (HLC) of the consumer ($H$) is shown in
Fig.~\ref{f:NS}(f).

With an increase in $\epsilon$ and choosing appropriate initial
conditions, Fig.~\ref{f:NS}(b) shows inhomogeneous limit cycles for
the same dispersal rate $d=0.35$. Corresponding trajectories are shown
in Fig.~\ref{f:NS}(g).  From this, it is clear that consumer species
in 16 patches are in-phase synchronized.  As like two patches, network
of 16 patches also show AD and OD.  Depending on initial conditions,
we get either IHLC or oscillation death.  Spatiotemporal dynamics and
time series of multi-clustered OD states are shown in
Figs.~\ref{f:NS}(c) and \ref{f:NS}(h), respectively.

For low dispersal rate ($d=0.1$), AD occurs.  The suppression of
oscillation in homogeneous steady states is shown in
Fig.~\ref{f:NS}(d) with the time series shown in Fig.~\ref{f:NS}(i).
Occurrence of rhythmogeneis is shown in Fig.~\ref{f:NS}(e) and
\ref{f:NS}(j) for $m=0.3$, $\epsilon=0.15$ and $d=0.3$.  The
environmentally coupled system of a network shows similar kind of
dynamics shown for 2 patches.  From the collective behavior shown in
Fig.~\ref{f:NS}, it is clear that HLC, IHLC, AD, OD and rhythmogeneis
all are also valid in a network connected by common dynamic
environment.

\section{Discussion}
\label{S:5}

In summary, in this paper we have modeled an ecological system
connecting local habitats through dispersal in a common environment in
which the considered environment is nonlinear and dynamic; at the same
time we also consider the interaction between the patches and the
common dynamic environment is essentially nonlinear, which is
controlled by ecologically relevant parameters like conversion
efficiency and half saturation constant.  We highlight that the
consumer interactions in three distinct ways such as the interaction
within the patch, between the patch and through the environment,
however all are interrelated in the collective behavior once coupling
involves in the system.  Specifically, our study reveals the role of
environmental coupling and the effects of inter-patch dispersal on a
spatially extended Rosenzweig--MacArthur model using various
bifurcation techniques.  With simple diffusive coupling in species
dispersal, the coupled system exhibits many interesting dynamics with
variability in spatial and environmental parameters.  Starting from
generation of oscillations, various synchronization processes in a
smooth oscillator and its suppression to different steady states such
as AD, OD and AD-OD transition are shown in an ecological system.
From different steady states and oscillation, the relationship between
synchrony and stability induced by dispersal as well as induced by the
dynamic environment has been identified.  Even though resource density
is not directly involved in dispersal, the intrinsic dynamics and the
consumer-resource interaction within the patch enables same
qualitative behavior in all the species involved in system.  Further,
the environmental correlation due to species dispersal enables the
persistence of the ecosystem via stabilization of oscillating
populations.

Earlier, landscape and metapopulation models have been used to predict
the species extinction risk and spatial pattern on ecological
processes \cite{Gu02,AkRa04,Wim06}.  Although our model depicts as
metapopulation model, but explicitly describes various ecological
perspectives qualitatively and quantitatively.  In particular,
dispersal enhances the occurrences of synchronized oscillations and
synchrony-stability relationship.  In biological system functioning,
synchronization of rhythms plays important role
\cite{GlaMc_book,Gold_book}.  Rhythmogenesis is an emergent behavior
of dynamical system since restoration of oscillations can be achieved
easily even if the system is in a steady state.  Also, due to the
presence of noise in ecological systems, oscillating species are prone
to extinction easily, so it is better to understand the factors which
enhance the synchrony--stability relationship.  Although many internal
and external factors influence in dispersal for successful
colonization in the new habitat, but different environmental
conditions due to heterogeneity enable the appearance and
disappearance of oscillations and transition of homogeneous steady
states to inhomogeneous steady states in this system.  In general,
spatial and environmental heterogeneity clearly distinguish the
synchrony--stability relationship induced by both dispersal and a
common dynamic environment.

Further, in the context of food web dynamics, food web complexity and
species movement pattern contribute largely to enrich the current
knowledge \cite{Amar08}.  Generally, active as well as passive
dispersal take place in natural systems.  Specifically, in active
dispersal, species are directly involved in movement whereas in
passive dispersal, species are being moved by other factors.  In this
work, we set active dispersal in consumer populations only, but in
addition, dispersal also happens in resource populations either
directly or indirectly \cite{LeKo04,Amar08}.  Further, time scale of
dispersal is slow as compare to temporal dynamics within the patch and
interconnected habitats are heterogeneous with various network
structure. Thus, further study is required to focus on different kind
of dispersal with different time scales in heterogeneous environments.

\begin{acknowledgments}
P.S.D. acknowledges financial support from SERB, Department of Science
and Technology (DST), India [Grant No.: YSS/2014/000057].
T.B. acknowledges the financial support from SERB, Department of
Science and Technology (DST), Govt. of India [Grant No.:
  SB/FTP/PS-005/2013].
\end{acknowledgments}

%

\end{document}